\documentclass{pnastwo}

\usepackage[dvips]{graphicx}

\usepackage{amssymb,amsfonts,amsmath}

\contributor{Submitted to Proceedings
of the National Academy of Sciences of the United States of America}
\url{www.pnas.org/cgi/doi/10.1073/pnas.0709640104}
\copyrightyear{2008}
\issuedate{Issue Date}
\volume{Volume}
\issuenumber{Issue Number}
\begin{document}

%%%%%%%%%%%%%%%%%%%%%%%%%%%%%%

%% For titles, only capitalize the first letter
%% \title{Almost sharp fronts for the surface quasi-geostrophic equation}

\title{Long range order and two-fluid behavior in heavy electron materials}

%% Enter authors via the \author command.
%% Use \affil to define affiliations.
%% (Leave no spaces between author name and \affil command)

%% Note that the \thanks{} command has been disabled in favor of
%% a generic, reserved space for PNAS publication footnotes.

%% \author{<author name>
%% \affil{<number>}{<Institution>}} One number for each institution.
%% The same number should be used for authors that
%% are affiliated with the same institution, after the first time
%% only the number is needed, ie, \affil{number}{text}, \affil{number}{}
%% Then, before last author ...
%% \and
%% \author{<author name>
%% \affil{<number>}{}}

%% For example, assuming Garcia and Sonnery are both affiliated with
%% Universidad de Murcia:
%% \author{Roberta Graff\affil{1}{University of Cambridge, Cambridge,
%% United Kingdom},
%% Javier de Ruiz Garcia\affil{2}{Universidad de Murcia, Bioquimica y Biologia
%% Molecular, Murcia, Spain}, \and Franklin Sonnery\affil{2}{}}

\author{K. R. Shirer\affil{1}{Department of Physics, University of California, Davis, CA 95616, USA},
A. C. Shockley\affil{1}{},
A. P. Dioguardi\affil{1}{},
J. Crocker\affil{1}{},
C.-H. Lin\affil{1}{},
N. apRoberts-Warren\affil{1}{},
D. M. Nisson\affil{1}{},
P. Klavins\affil{1}{},
J. C. Cooley\affil{2}{Los Alamos National Laboratory, Los Alamos, New Mexico 87545, USA},
Y.-F. Yang\affil{3}{Beijing National Laboratory for Condensed Matter Physics and Institute of Physics, Chinese Academy of Sciences, Beijing 100190 China}
\and
N. J. Curro\affil{1}{}}

\contributor{Submitted to Proceedings of the National Academy of Sciences
of the United States of America}

%% The \maketitle command is necessary to build the title page.
\maketitle

%%%%%%%%%%%%%%%%%%%%%%%%%%%%%%%%%%%%%%%%%%%%%%%%%%%%%%%%%%%%%%%%
\begin{article}

\begin{abstract}

The heavy electron Kondo liquid is an emergent state of condensed matter that displays universal behavior independent of material details.  Properties of the heavy electron liquid are best probed by NMR Knight shift measurements, which provide a direct measure of the behavior of the heavy electron liquid that emerges below the Kondo lattice coherence temperature as the lattice of local moments hybridizes with the background conduction electrons.  Because the transfer of spectral weight between the localized and itinerant electronic degrees of freedom is gradual, the Kondo liquid typically coexists with the local moment component until the material orders at low temperatures.  The two-fluid formula captures this behavior in a broad range of materials in the paramagnetic state.   In order to investigate two-fluid behavior and the onset and physical origin of different long range ordered ground states in heavy electron materials, we have extended Knight shift measurements to URu$_2$Si$_2$, CeIrIn$_5$ and CeRhIn$_5$. In CeRhIn$_5$ we find that the antiferromagnetic order is preceded by a relocalization of the Kondo liquid, providing independent evidence for a local moment origin of antiferromagnetism. In URu$_2$Si$_2$ the hidden order is shown to emerge directly from the Kondo liquid and so is not associated with local moment physics. Our results imply that the nature of the ground state is strongly coupled with the hybridization in the Kondo lattice in agreement with phase diagram proposed by Yang and Pines.

 \end{abstract}

%% When adding keywords, separate each term with a straight line: |

%% Optional for entering abbreviations, separate the abbreviation from
%% its definition with a comma, separate each pair with a semicolon:
%% for example:
%% \abbreviations{SAM, self-assembled monolayer; OTS,
%% octadecyltrichlorosilane}

% \abbreviations{}

%% The first letter of the article should be drop cap: \dropcap{}
%\dropcap{I}n this article we study the evolution of ''almost-sharp'' fronts

%% Enter the text of your article beginning here and ending before
%% \begin{acknowledgements}
%% Section head commands for your reference:
%% \section{}
%% \subsection{}
%% \subsubsection{}

\dropcap{C}ompetition between different energy scales gives rise to a rich spectrum of emergent ground states in strongly correlated electron materials. In the heavy fermion compounds, a lattice of nearly localized f-electrons interacts with a sea of conduction electrons, and depending on the magnitude of this interaction different types of long range order may develop at low temperatures \cite{LohneysenQPTreview}. The Kondo lattice model strives to capture the essential physics of heavy fermion materials by considering the various magnetic interactions between the conduction electron spins, $S_c$, and the local moment spins, $S_f$ \cite{doniach}.  Different ground states can emerge depending on the relative strengths of the interaction, $J$, between $S_c$ and $S_f$ and the intersite interaction, $J_{ff}$, between the $S_f$ spins \cite{YangPinesNature,YangPinesPNASdraft}.  Much of the physics of the phase diagram is driven by a quantum critical point, which separates long-range ordered ground states from those in which  the local moments have fully hybridized with the conduction electrons to form itinerant states with large effective masses and  large Fermi surfaces \cite{YamamotoSiPRL2007}. In several materials quantum critical fluctuations give rise to anomalous non-Fermi liquid behavior in various bulk transport and thermodynamic quantities \cite{YRSnature,stockertNS,TusonPNAS}.

In recent years, evidence has emerged that in the high temperature disordered phase the electronic degrees of freedom simultaneously exhibit both itinerant and localized behavior \cite{NPF}. Below a temperature $T^*$ that marks the onset of hybridization or lattice coherence, several experimental quantities exhibit a temperature dependence that is well described by a fluid of hybridized heavy quasiparticles coexisting with partially screened local moments \cite{YangPinesNature,YangDavidPRL}. In the two fluid picture, the transfer of spectral weight from the local moments to the heavy electrons is described by a quantity $f(T)$ \cite{YangDavidPRL}. Above a temperature $T^*$, the local moments and conduction electrons remain uncoupled. Below $T^*$ the local moments gradually dissolve into the hybridized heavy electron fluid as the temperature is lowered. This crossover can be measured directly via nuclear magnetic resonance (NMR) Knight shift experiments \cite{CurroKSA}. The contribution to the Knight shift from the heavy electrons, $K_{HF}$, exhibits a universal logarithmic divergence with decreasing temperature below $T^*$ in the paramagnetic state. In superconducting CeCoIn$_5$ this scaling persists down to $T_c$ indicating that the condensate emerges from the heavy electron degrees of freedom \cite{Yang2009}. Relatively little is known, however, about how this scaling behaves in materials with other types of ordered ground states. Here we report Knight shift data in both URu$_2$Si$_2$ and CeRhIn$_5$. In the former the scaling persists down to the onset of hidden order at $T_{HO}$; but in the latter the heavy electron spectral weight reverses course and the local moments relocalize above the Neel temperature, $T_N$, similar to the antiferromagnet CePt$_2$In$_7$ \cite{CePt2In7KnightShift}. These results imply generally that the heavy electron fluid is either unstable to a Fermi surface instability such as hidden order or superconductivity, or collapses to form ordered local moments.

\section{Knight Shift anomalies} The Zeeman interaction of an isolated nuclear spin $\mathbf{\hat{I}}$ in an external magnetic field $\mathbf{H}_0$ is given by $\mathcal{H}_Z = \gamma\hbar\mathbf{\hat{I}}\cdot\mathbf{H}_0$, where $\gamma$ is the gyromagnetic ratio and $\hbar$ is the reduced Planck constant.  In this case the nuclear spin resonance frequency is given by the Larmor frequency $\omega_L = \gamma H_0$.  In condensed matter systems the nuclear spins typically experience a hyperfine coupling to the electron spins, $\mathcal{H}_{hyp} = \gamma\hbar g\mu_B\mathbf{\hat{I}}\cdot\mathbf{A}\cdot\mathbf{S}$, where $\mathbf{S}$ is the electron spin and $\mathbf{A}$ is the hyperfine coupling. The hyperfine coupling is in general a tensor that depends on the details of the bonding and quantum chemistry of the material of interest.  $\mathbf{S}$ can be the  unpaired spin moment of an unfilled shell, the net conduction electron spin of a Fermi sea polarized by a magnetic field, or the spin-orbit coupled moment of a localized electron.  $\mathbf{S}$ may or may not be located on the same ion as the nuclear moment; the former is referred to as \emph{on-site} coupling and the latter is  \emph{transferred} coupling.  In the paramagnetic state the electron spin is polarized by the external field $\mathbf{S} = \mathbf{\chi}\cdot\mathbf{H}_0/g\mu_B$, where $\mathbf{\chi}$ is the susceptibility.  Thus the nuclear spin experiences an effective Hamiltonian:
\begin{equation}
\mathcal{H}_{Z}+\mathcal{H}_{hyp} = \gamma\hbar\mathbf{\hat{I}}\cdot\left(\mathbf{1} + \mathbf{K}\right)\cdot\mathbf{H}_0,
\end{equation}
where the Knight shift tensor $\mathbf{K} = \mathbf{A}\cdot\mathbf{\chi}$ is unit-less.  The resonance frequency is given by $\omega = \omega_L(1+K(\theta,\phi))$, where $K(\theta,\phi) = \mathbf{H}_0\cdot\mathbf{A}\cdot\mathbf{\chi}\cdot\mathbf{H}_0/H_0^2$ depends on the polar angles $\theta$ and $\phi$ that describe the relative orientation of the field with respect to the principal axes of the Knight shift tensor.  For the isotropic case the shift $K = A\chi$ is independent of direction.  Strictly speaking the \emph{Knight} shift was originally defined as the paramagnetic shift from the Pauli susceptibility in metals, but in practice refers to any shift $K$ arising from the electronic susceptibility.

The magnetic susceptibility and the Knight shift can be measured independently. One can extract the hyperfine coupling $A$ by plotting the Knight shift versus the susceptibility with temperature as an implicit parameter (such a plot is traditionally known as a \emph{Clogston-Jaccarino} plot \cite{PtNMR}).  If there is a single electronic spin component that gives rise to the magnetic susceptibility, then the Clogston-Jaccarino plot is a straight line with slope $A$ and intercept $K_0$.  $K_0$ is a temperature independent offset that is usually given by the orbital susceptibility and diamagnetic contributions \cite{abragambook}.

 \begin{figure}
\centering
\includegraphics[width=\linewidth]{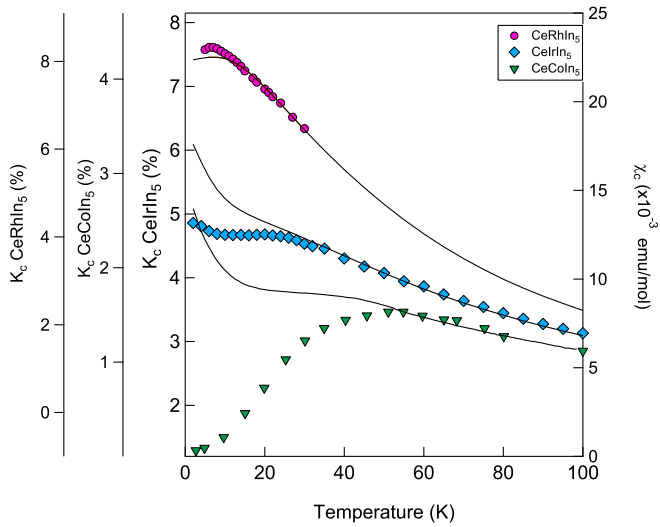}
\caption{\label{fig:115comparison}The Knight shift (solid points) and the bulk susceptibility (solid lines) of the In(1) for field along the $c-$direction in CeMIn$_5$ for M = Rh, Ir and Co. $T^*$ is the temperature where the two quantities no longer are proportional to one another, and is clearly material dependent. In some cases $K$ appears to exceed $\chi$ and in others $K$ decreases.}
\end{figure}

\subsection{Breakdown of linearity and failure of local pictures} All known heavy fermion materials exhibit a Knight shift anomaly where the Clogston-Jaccarino plot deviates  from linearity below an onset temperature, $T^*$ \cite{CurroKSA}. This phenomenon is also visible in a plot comparing $K$ and $\chi$ versus temperature, as seen in Fig. \ref{fig:115comparison}, where $T^*$ marks the temperature below which $K$ no longer tracks $\chi$.  This breakdown can arise from the presence of impurity phases, which contribute to the bulk susceptibility but not to the Knight shift.  However Knight shift anomalies occur ubiquitously in single phase highly pure heavy fermion materials and are intrinsic phenomena.  Several different theories have been proposed to explain these anomalies.  These theories generally fall into two categories: (i) a temperature dependent hyperfine coupling $A(T)$, or (ii) multiple spin degrees of freedom.  It has been argued that $A$ acquires a temperature dependence either because of Kondo screening \cite{coxKSA} or because of different populations of crystal field levels of the 4f(5f) electrons in these materials \cite{YasuokaCeCu2Si2}.  A major flaw in such an argument is that the hyperfine coupling is determined by large energy scales involving exchange integrals between different atomic orbitals, and since the Kondo and/or crystal field interactions are several orders of magnitude smaller they cannot give rise to large perturbations in the hyperfine couplings \cite{MilaRiceHamiltonian}. Furthermore, in materials such as the CeMIn$_5$ system both the Knight shift anomaly and the crystalline electric field interaction are well characterized and are clearly uncorrelated (see Table \ref{tab:CEF115}) \cite{CEF115study}.

\begin{figure}
\centering
\includegraphics[width=\linewidth]{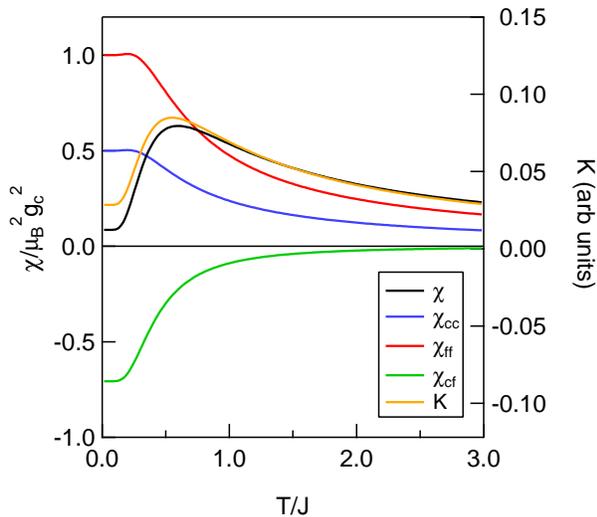}
\caption{\label{fig:chicomponents}The components of the magnetic susceptibility for the
model spin system.  The Knight shift begins to deviate from
 the bulk susceptibility below a temperature $T\approx J$.
 In this case, we have used $g_f/g_c=\sqrt{2}$, and $B/A = 1.5$.
 If $g_f = g_c$, then $\chi(T\rightarrow 0) \rightarrow 0$, as expected for a singlet ground state.}
\end{figure}

\subsection{Two spin components} Since Kondo lattice materials have both localized f-electrons and itinerant conduction electrons, a theory which includes different hyperfine couplings to multiple spin degrees of freedom fits better. The two-fluid picture correctly predicts the temperature dependence of the Knight shift anomalies seen experimentally in heavy fermion materials. In the Kondo lattice model the local moment  spins, $\mathbf{S}_f$ interact with a sea of conduction electron spins, $\mathbf{S}_c$ through a Kondo interaction:
\begin{equation}
\mathcal{H}_{K} = J \sum_i \mathbf{S}_f(\mathbf{r}_i)\cdot\mathbf{S}_c + \sum_{ij}J_{ff}(\mathbf{r}_{ij})\mathbf{S}_f(\mathbf{r}_i)\cdot\mathbf{S}_f(\mathbf{r}_j)
\end{equation}
where the sums are over the lattice positions $\mathbf{r}_i$ of the localized f electrons.  The general solution of the Kondo lattice has not been resolved to date, but the two-fluid description given by Nakatsuji, Fisk, Pines and Yang provides a phenomenological framework to describe the phase diagram of the Kondo lattice problem \cite{YangPinesNature,YangPinesPNASdraft,NPF,YangDavidPRL}.  This approach offers a natural interpretation of the NMR Knight shift anomaly in terms of two hyperfine couplings to the two different electron spins:
\begin{equation}
\mathcal{H}_{hyp}=\gamma\hbar g\mu_B\hat{\mathbf{I}}\cdot(A \mathbf{S}_c+B\mathbf{S}_f)
\end{equation}
where $A$ and $B$ are the hyperfine couplings to the conduction electron and local moment spins, respectively \cite{CurroKSA}.  The susceptibility is the sum of  three contributions $\chi =\chi_{cc}+ 2 \chi_{cf} + \chi_{ff}$, where $\chi_{\alpha\beta} = \langle\mathbf{S}_{\alpha}\mathbf{S}_{\beta}\rangle$ with $\alpha,\beta = c,f$. In the paramagnetic state $\langle S_c\rangle = \left(\chi_{cc} + \chi_{cf}\right)H_0$ and $\langle S_f\rangle = \left(\chi_{ff} + \chi_{cf}\right)H_0$, thus the Knight shift is given by:
\begin{equation}
K =K_0+ A\chi_{cc} + (A+B)\chi_{cf} + B\chi_{ff}.
\label{eqn:KnightShift}
\end{equation}
The Knight shift weighs the different correlation functions separately than the total susceptibility, which leads to important consequences if the correlation functions have different temperature dependences.  $K \sim \chi$ if and only if $A=B$.  If $A\neq B$, then the Clogston-Jaccarino plot will no longer be linear.  By measuring $K$ and $\chi$ independently, one can exploit this difference to extract linear combinations of these individual susceptibilities.  No other experimental technique can access these susceptibilities.

\begin{figure}
\centering
\includegraphics[width = \linewidth]{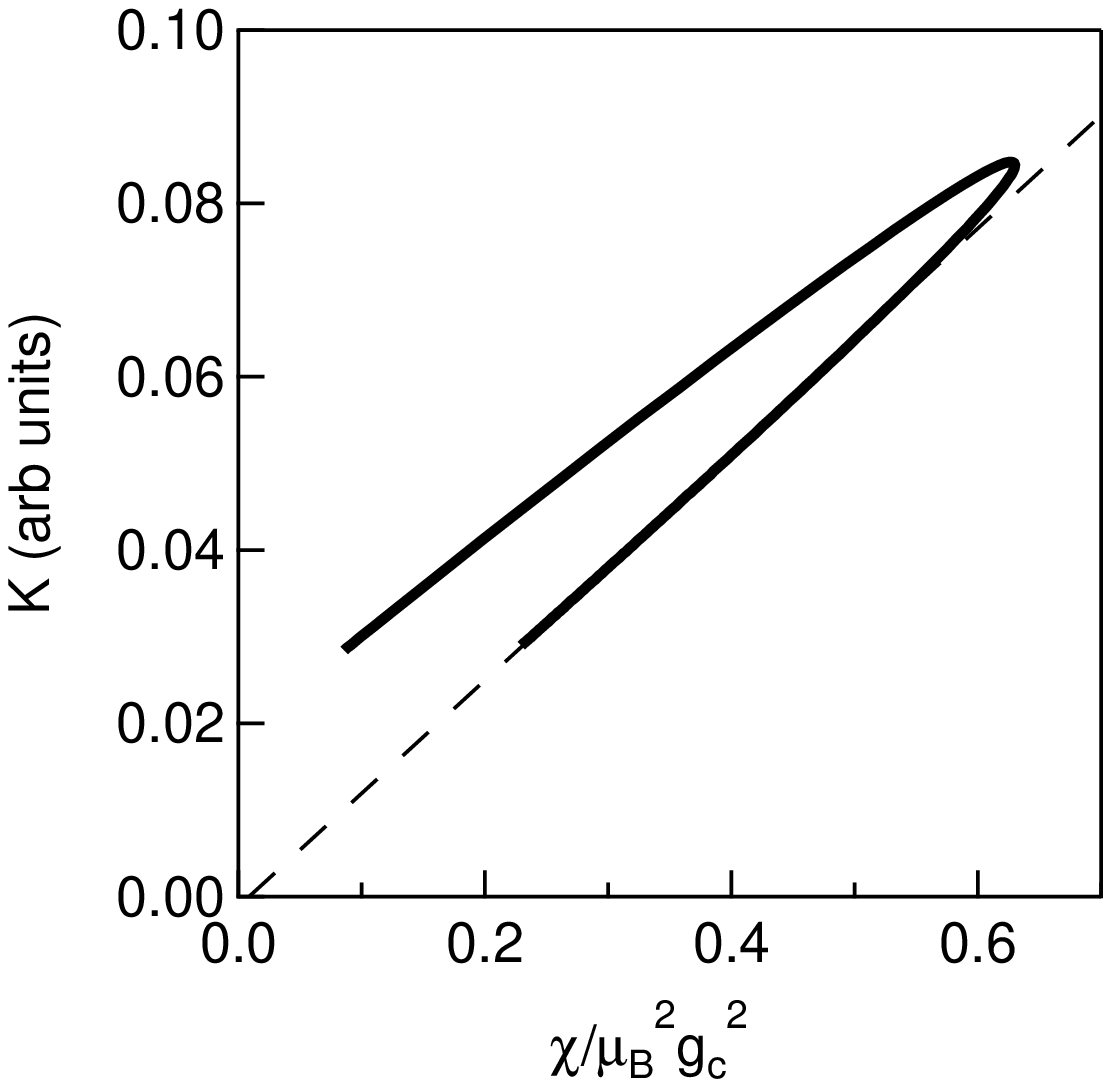}
\caption{\label{fig:KvsChiModel} Knight shift versus the bulk susceptibility for the model spin system with temperature as an implicit parameter.  The Knight shift begins to deviate from the bulk susceptibility
below a temperature $T\approx J/k_B$. In this case, we have used $g_f/g_c=\sqrt{2}$, and $B/A = 1.5$.}
\end{figure}

 \begin{figure}
\centering
\includegraphics[width=\linewidth]{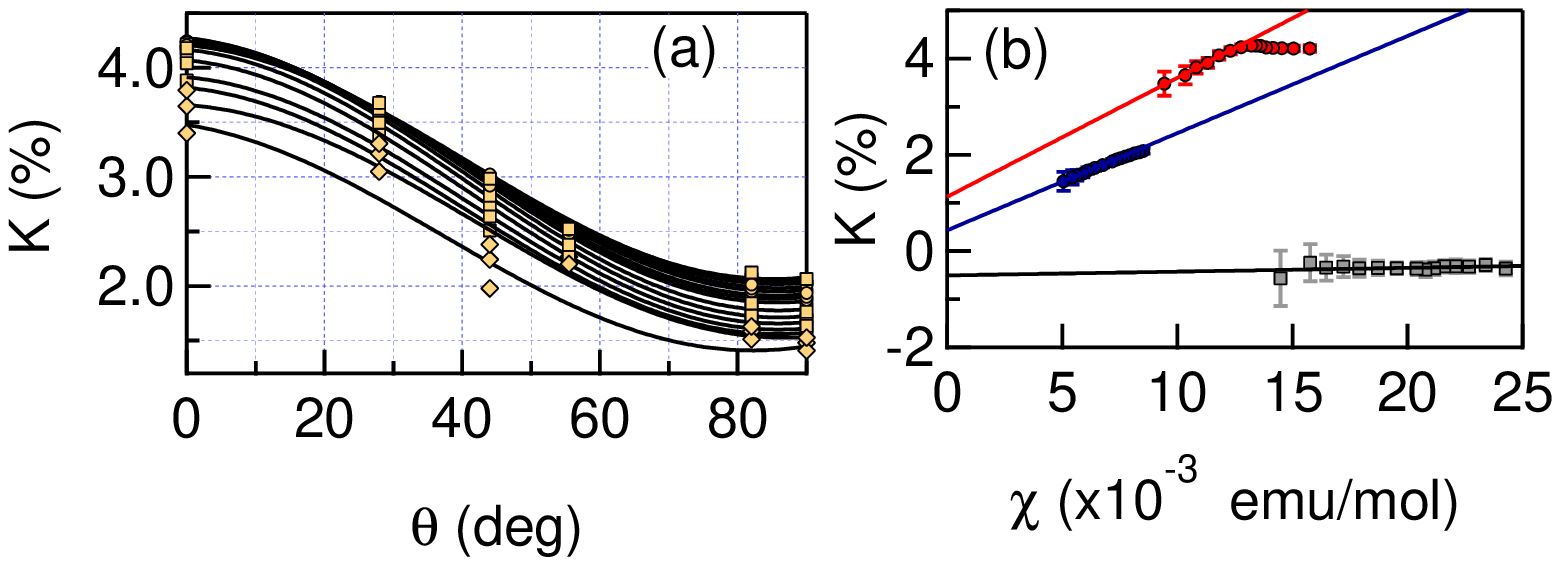}
\caption{\label{fig:Ir115Kdata}(a) The Knight shift of the In(1) in CeIrIn$_5$ versus angle at several different temperatures.  Solid points are fits to Eq. \ref{eqn:KnightShift}. (b) $K_{aa}$ (blue), $K_{cc}$ (red) and $K_{ac}$ (gray) versus $\chi_{aa}$, $\chi_{cc}$ and $\chi_{aa} + \chi_{cc}$. Solid lines are linear fits as described in the text.}
\end{figure}

\subsection{Free spin model}  The general solution for the susceptibilities $\chi_{\alpha\beta}$ of the Kondo lattice problem has not been solved to date. It is useful, however, to consider a simplified model.  Consider the case of two free $S=\frac{1}{2}$ spins $\mathbf{S}_c$ and $\mathbf{S}_f$ that experience a Heisenberg coupling $J$.  The Hamiltonian for this spin system in the presence of an external magnetic field $\mathbf{H}_0$ is given by:
\begin{equation}
\hat{\mathcal{H}} = g_c\mu_B \mathbf{S_c}\cdot\mathbf{H}_0 + g_f\mu_B \mathbf{S_f}\cdot\mathbf{H}_0 + J\mathbf{S}_c\cdot\mathbf{S}_f,
\label{eqn:coupledfreespins}
\end{equation}
where $\mu_B$ is the Bohr magneton, and $g_{c,f}$  are the g-factors of the two different spins.  This simplified model is clearly an \emph{incorrect}  physical description of the Kondo lattice problem where the spins are not free but part of a complex many-body problem involving the Fermi sea and various intersite interactions. This model is the simplest case, which captures the essence of the Knight shift anomaly and can be solved exactly.

The free energy is given by $F= -k_B T \ln Z$, where $Z$ is the partition function; the susceptibility of the system is given by the second derivative of the free energy with respect to field: $\chi = \frac{\partial^2 F}{\partial H^2}|_{H\rightarrow 0}$.  The exact susceptibility is given as the sum of three contributions:
\begin{eqnarray}
\chi_{cc}(x) &=& \left(\frac{g_c^2\mu_B^2}{4 k_B T}\right) \frac{2 \left(e^{x}+x-1\right) }{\left(3+e^{x}\right) x}    \\
\chi_{ff}(x) &=& \left(\frac{g_f^2\mu_B^2}{4 k_B T}\right) \frac{2 \left(e^{x}+x-1\right) }{\left(3+e^{x}\right) x}   \\
\chi_{cf}(x) &=&  -\left(\frac{g_c g_f\mu_B^2}{4 k_B J}\right) \frac{2\left(e^{x}-x -1\right)}{3+e^{x}}
\end{eqnarray}
where $x = J/k_BT$. The temperature dependence of these three susceptibilities is shown in Fig. \ref{fig:chicomponents}. In the high temperature limit ($T\gg J/k_B$) $\chi_{cc}$ and $\chi_{ff}$ exhibit Curie behavior and $\chi_{cf} \rightarrow 0$ as expected.  For low temperature ($T\lesssim J/k_B$) the behavior is modified by the tendency to form a singlet ground state. As shown in Fig. \ref{fig:chicomponents} $\chi_{cc}$, $\chi_{ff}$ and $\chi_{cf}$ change dramatically below $T\approx J/k_B$ and saturate at low temperatures. The Knight shift, given by Eq. \ref{eqn:KnightShift}, is shown as well in Fig. \ref{fig:chicomponents}. Clearly $K$ tracks $\chi$ for $T \gg J/k_B$ but deviates below $T \approx J/k_B$, creating an anomaly in the Clogston-Jaccarino plot in Fig. \ref{fig:KvsChiModel}. The breakdown of linearity arises because $|\chi_{cf}|$ increases below $T\approx J/k_B$ and has a \emph{different temperature dependence} than $\chi_{cc}$ and $\chi_{ff}$.

 \begin{figure}
\centering
\includegraphics[width=\linewidth]{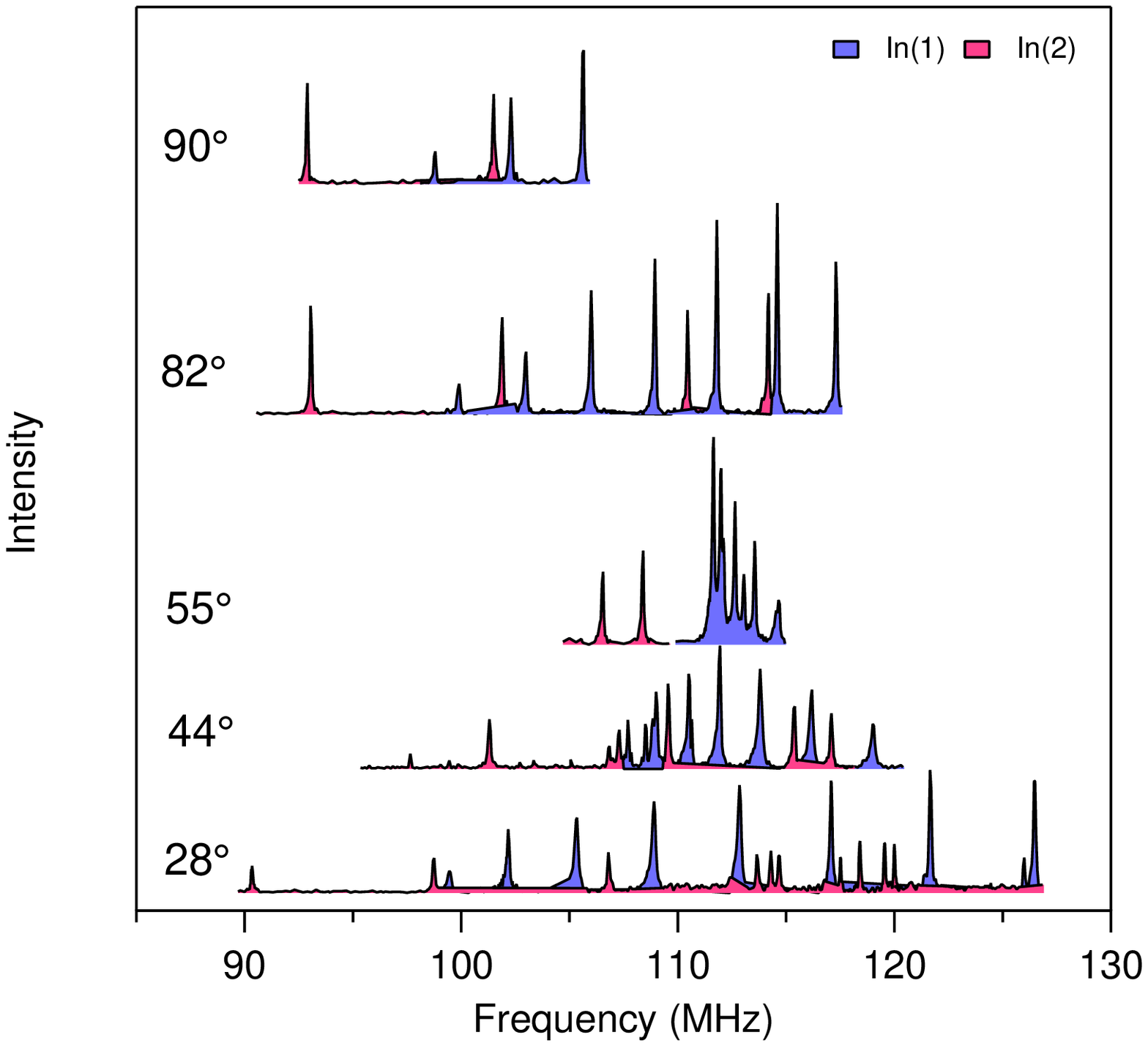}
\caption{\label{fig:Ir115spectra} Spectra of the In(1) and In(2) in CeIrIn$_5$ at 6 K and constant field of 11.7 T, where $\theta$ is the angle between the $c$ axis and the field. There are multiple satellite transitions for each site due to the quadrupolar splitting \cite{CurroHFreview}.}
\end{figure}

\subsection{Two-fluid model} The three individual susceptibilities $\chi_{\alpha\beta}$  will exhibit different temperature dependences in a Kondo lattice than the free spin model considered above. In particular, at high temperatures $\chi_{cc}$ is given by the temperature independent Pauli susceptibility of the conduction electrons, and $\chi_{ff}$ is given by a Curie-Weiss susceptibility of the local moments. Therefore above a crossover temperature, $T^*$, $\chi_{ff}$ dominates and $K=K_0+B\chi$. Below $T^*$ $\chi_{cf}$ becomes significant and the growth of this component can be measured by the quantity $K_{HF} = K-K_0 - B\chi = (A-B)(\chi_{cf} + \chi_{cc})$.  The two-fluid model postulates that $K_{HF}$ is proportional to the susceptibility of the heavy electron fluid; its temperature dependence probes both growth of hybridization and its relative spectral weight \cite{YangPinesNature,YangDavidPRL}.
Empirically it has been found that in the paramagnetic phase of a broad range of materials, $K_{HF}(T)$ varies as:
\begin{equation}
K_{HF}(T) = K_{HF}^0(1-T/T^*)^{3/2}[1+\log(T^*\!/T)]
\label{eqn:YangPines}
\end{equation}
where $K_{HF}^0$ is a constant \cite{YangDavidPRL}. In other words, $K_{HF}$ exhibits a universal scaling with the quantity $T/T^*$, where $T^*$ and $K_{HF}^0$ are material dependent quantities.  $K_{HF}^0$ is proportional to $A-B$ and can be either positive or negative. $T^*$ agrees well with several other experimental measurements of the coherence temperature of the Kondo lattice \cite{YangDavidPRL}. The Knight shift thus provides a direct probe of the susceptibility of the emergent heavy electron fluid in the paramagnetic state.   There are, however, several pieces of missing information regarding the behavior of the heavy electron component that need to be addressed. For example, the Knight shift anomaly is strongly anisotropic in some materials, and is not been clear whether this is an intrinsic feature of the heavy electron fluid or rather a reflection of the anisotropy of the hyperfine couplings \cite{CurroKSA}.  Furthermore,  the Yang-Pines scaling works remarkably well for temperatures down to $\sim 10$\% of $T^*$, but until recently there has been relatively little information about the evolution of the heavy electron fluid at lower temperature particularly when long-range order develops. We address both of these issues in turn below.

 \begin{figure}
\centering
\includegraphics[width=\linewidth]{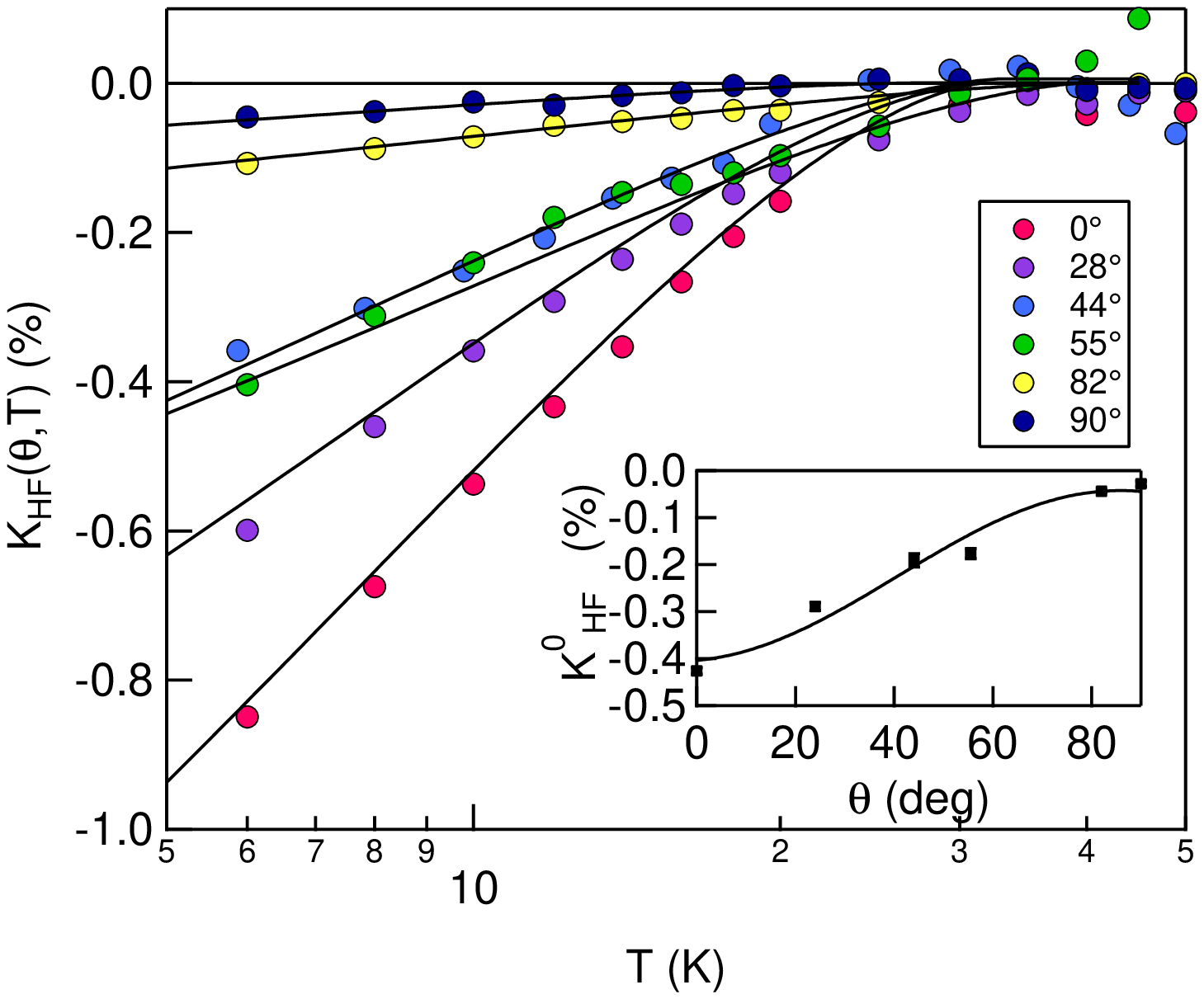}
\caption{\label{fig:Ir115Khf} The heavy electron component of the shift in CeIrIn$_5$ for the In(1) as a function of angle and temperature. The solid lines are fits to Eq. \ref{eqn:YangPines}.  Inset: The fitted value $K^0_{HF}$ versus angle.  The solid line fit, described in the text, indicates the tensor nature of the hyperfine coupling $\mathbf{A}$.}
\end{figure}

 \begin{figure}
\centering
\includegraphics[width=\linewidth]{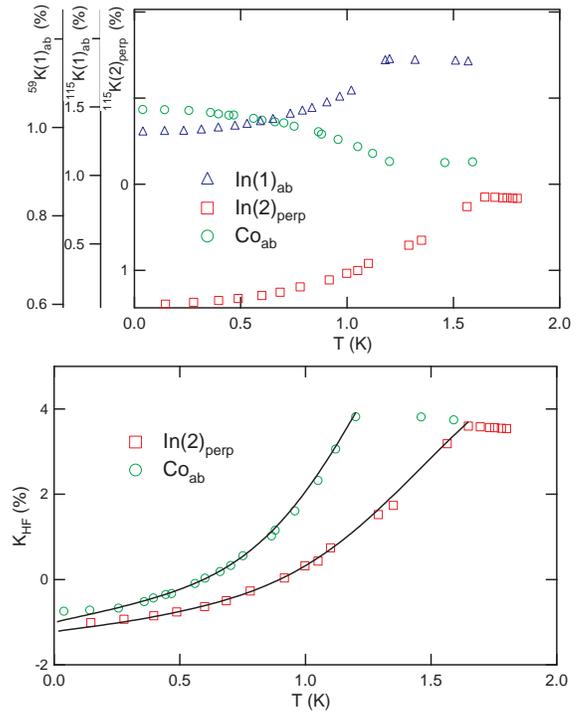}
\caption{\label{fig:Co115KS} (upper panel) Knight shift of the In(1), In(2) and Co in CeCoIn$_5$ in the paramagnetic state for field in the $ab$ plane.  For In(2) there are two distinct crystallographic positions depending on the direction of the field. (lower panel) The heavy electron susceptibility $K_{HF}$ in the superconducting state. The solid lines are calculations as discussed in the text.  Data are reproduced from \cite{Yang2009}.}
\end{figure}

\begin{figure}
\centering
\includegraphics[width=0.8\linewidth]{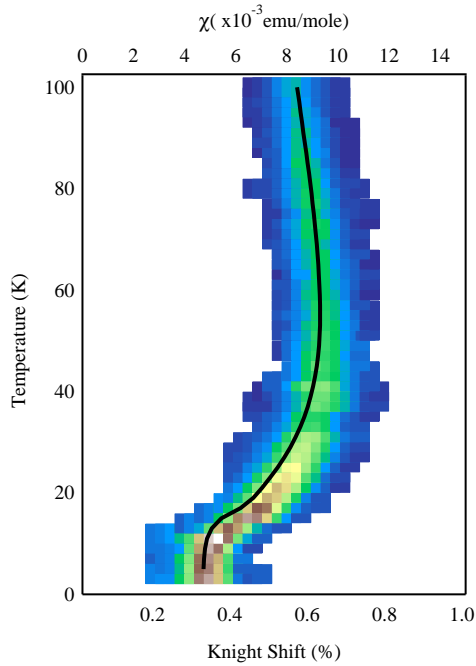}
\caption{\label{fig:URSmap} Spectra of the $^{29}$Si in URu$_2$Si$_2$ as a function of temperature. Color intensity (arb units) corresponds to the integral of the NMR spin echo. The solid black line is the bulk susceptibility.}
\end{figure}

\begin{figure}
\centering
\includegraphics[width=1.0\linewidth]{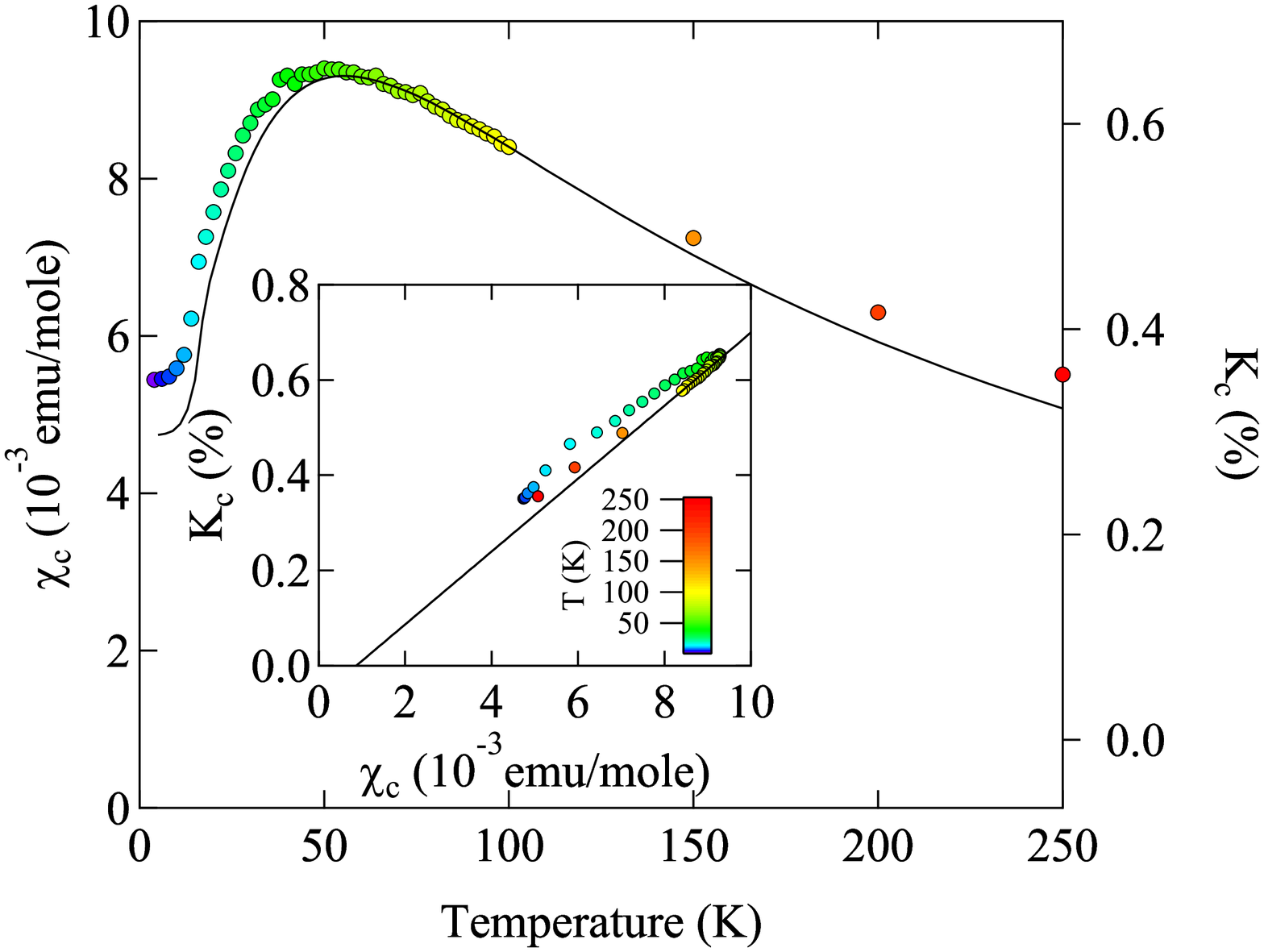}
\caption{\label{fig:ShiftURS} The Knight shift of the Si (colored circles) and the susceptibility (solid line) in URu$_2$Si$_2$ for field along the $c$ direction. The inset shows $K$ versus $\chi$.}
\end{figure}

\begin{figure}
\centering
\includegraphics[width=1.0\linewidth]{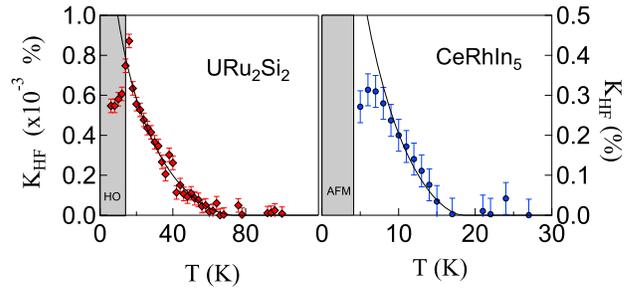}
\caption{\label{fig:KhfvsT} $K_{HF}(T)$  versus $T$  for  URu$_2$Si$_2$ (left)  and CeRhIn$_5$ (right).  The solid black line are best fits to Eq. \ref{eqn:YangPines}. In URu$_2$Si$_2$ $K_{HF}$ grows until the sudden onset of hidden order, whereas in CeRhIn$_5$ $K_{HF}$ deviates from the scaling form below 8 K indicating the relocalization of the local moments.}
\end{figure}

\subsection{Anisotropy} One of the striking features of the emergence of the heavy electron component in Kondo lattice materials is the anisotropy of the Knight shift anomaly.  For example, in CeCoIn$_5$ the Knight shift of the In(1) site exhibits a strong anomaly for $\mathbf{H}_0\parallel c$ but no anomaly for $\mathbf{H}_0\perp c$ \cite{CurroAnomalousShift}.  In order to probe this behavior in greater detail, we have measured the full Knight shift tensor in CeIrIn$_5$.  CeIrIn$_5$ is isostructural to CeCoIn$_5$, and is a superconductor with $T_c = 0.4$ K.  A single crystal grown in In flux was mounted in a custom-made goniometer probe, and the orientation of the $c-$axis of the crystal was varied from 0 to 90 degrees.  $^{115}$In has spin $I=\frac{9}{2}$, and there are two different crystallographic sites for In in this material; consequently there are several different transitions as seen in Fig. \ref{fig:Ir115spectra}. Spectra were obtained as a function of orientation and temperature, and the Knight shift was extracted after correcting for the quadrupolar shift of the resonance \cite{curroEMRquadSC}.  Fig. \ref{fig:Ir115Kdata} shows $K$ versus $\theta$, where $\theta$ is the angle between $c$ and $\mathbf{H}_0$.

 \begin{figure}
\centering
\includegraphics[width=\linewidth]{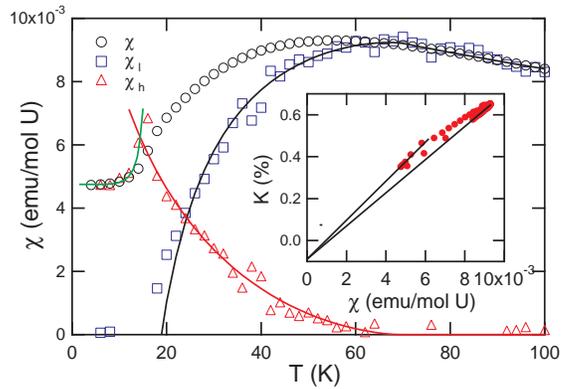}
\caption{\label{fig:URS}  Two-fluid analysis of the magnetic susceptibility in URu$_2$Si$_2$ by using the one-component behavior above $T^*$ and below $T_L$. The solid lines are fit to the scaling formula of the Kondo liquid and the mean-field formula of the hybridized spin liquid for $T>T_{HO}$ and the Yosida function for $T<T_{HO}$ following the BCS prediction.}
\end{figure}

\begin{figure}
\centering
\includegraphics[width=1.0\linewidth]{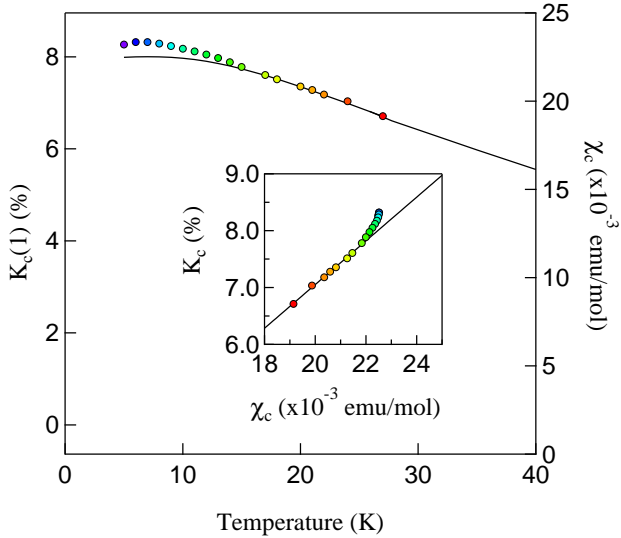}
\caption{\label{fig:ShiftRh115} The Knight shift of the In(1) (colored circles) and the susceptibility (solid line) in CeRhIn$_5$ for field along the $c$ direction. The inset shows $K$ versus $\chi$.}
\end{figure}

The strong angular dependence can be understood in terms of the tensor nature of the hyperfine coupling. For $T>T^*$  the tensor is given by:
\begin{equation}
\label{eqn:Btensor}
\mathbf{B} = \left(
               \begin{array}{ccc}
                 B_{aa} & B_{ab} & B_{ac} \\
                 B_{ab} & B_{bb} & B_{bc} \\
                 B_{ac} & B_{bc} & B_{cc} \\
               \end{array}
             \right)
\end{equation}
in the basis defined by the tetragonal unit cell defined by the unit vectors $a$ and $c$, and the susceptibility tensor is:
\begin{equation}
\mathbf{\chi} = \left(
               \begin{array}{ccc}
\chi_{aa} & 0 & 0 \\
0 & \chi_{aa} & 0 \\
0 & 0 & \chi_{cc} \\
               \end{array}
             \right).
\end{equation}
We therefore have:
\begin{equation}
K(\theta) = K_{aa}\sin^2\theta + K_{cc}\cos^2\theta + K_{ac}\sin\theta\cos\theta
\label{eqn:Ktheta}
\end{equation}
where $\theta$ is the angle between $\mathbf{H}_0$ and the $c-$axis, $K_{aa} = K_{aa}^0 + B_{aa}\chi_{aa}$, $K_{cc} = K_{cc}^0+ B_{cc}\chi_{cc}$, and $K_{ac} = K_{ac}^0 + B_{ac}(\chi_{aa} + \chi_{cc})$.  The solid lines in Fig. \ref{fig:Ir115Kdata}(a) are fits to this equation, and Fig. \ref{fig:Ir115Kdata}(b) shows  $K_{aa}$, $K_{cc}$ and $K_{ac}$ versus $\chi_{aa}$, $\chi_{cc}$ and $\chi_{aa} + \chi_{cc}$, respectively. The fitted values are given in Table \ref{tab:hypIr115}. The solid lines are fits that yield the hyperfine couplings $B_{aa}$, $B_{cc}$, $B_{ac}$ and the orbital shifts $K_{aa}^0$, $K_{cc}^0$ and $K_{ac}^0$.
The fact that $B_{ac} \approx 0$ within error bars implies that the principal axes of the hyperfine tensor coincide with those of the unit cell.  On the other hand, $K_{ac}^0 \neq 0$ implies that the orbital shift tensor is not diagonal in the crystal basis.  Rather it is rotated by an angle of $\sim 28^{\circ}$  from the $c-$direction, suggesting a possible role of directionality between the In 5p orbitals and the Ce 4f orbitals.

The emergence of the heavy fermion fluid is evident by computing $K_{HF}(\theta, T) = K(\theta,T) - K(\theta)$, where $K(\theta)$ is given by Eq. \ref{eqn:Ktheta} and using the fitted parameters.  This data in shown in Fig. \ref{fig:Ir115Khf}, as well as fits to Eq. \ref{eqn:YangPines}, which give a value of $T^* = 31 (5)$ K.  From this data, it is clear that $T^*$ does not vary as a function of angle, but rather the overall scale of the the quantity $K_{HF}^0$ decreases with angle until it reaches approximately zero by $\theta = 90^{\circ}$.  The angular dependence clearly indicates that $K_{HF}(\theta, T)$ does not simply vanish for the perpendicular direction, but  decreases gradually as a function of orientation in a continuous fashion.  The constant $K_{HF}^0$ is anisotropic, but $T^*$ is not. In fact, $\mathbf{K}_{HF}^0$ is proportional to $\mathbf{A} - \mathbf{B}$, and is therefore a tensor quantity itself, which is clearly seen in the angular dependence shown in the inset of Fig. \ref{fig:Ir115Khf}.  The solid line is a fit to $K_{HF}^0 = K_{HF,c}^0\cos^2\theta + K_{HF,a}^0\sin^2\theta$. It is likely that the anisotropy arises primarily because of the hyperfine couplings, but it is not possible to rule out an additional contribution to the anisotropy from the quantity $\chi_{cf}+ \chi_{cc}$.  The fact that $T^*$ does not depend on orientation suggests that the field $\mathbf{H}_0$ has little effect on the onset of coherence.

\begin{figure}
\centering
\includegraphics[width=1.0\linewidth]{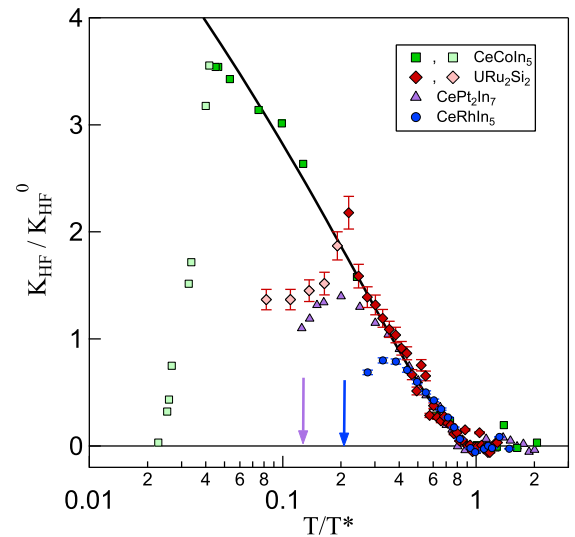}
\caption{\label{fig:scaling}  $K_{HF}(T)$ (normalized) versus $T/T^*$  for CeCoIn$_5$, CePt$_2$In$_7$, URu$_2$Si$_2$, and CeRhIn$_5$. Data points in the superconducting and hidden order states are lighter shades. The solid black line is Eq. \ref{eqn:YangPines}, and the vertical arrows indicate $T_N$ for CePt$_2$In$_7$ and CeRhIn$_5$. Data for the CeCoIn$_5$ and  CePt$_2$In$_7$ are reproduced from Refs. \cite{CePt2In7KnightShift} and \cite{Yang2009}.}
\end{figure}

\section{Superconducting order}

A key question in Kondo lattice materials is how the emergent heavy electron fluid and the remaining local moments interact to give rise to long range ordered states. Several materials exhibit a superconducting ground state, and it has long been assumed that the condensate forms from heavy electron quasiparticles. Recently we showed that this interpretation is supported by direct measurements of the Knight shift in the superconducting state \cite{Yang2009}.   Fig. \ref{fig:Co115KS} displays $K(T)$ and $K_{HF}(T)$ for the In and Co sites in CeCoIn$_5$.  Surprisingly, the total Knight shift $K(T)$ \emph{increases} for some of the sites below $T_c$, and decreases for others.  The normal expectation is that the spin susceptibility decreases below $T_c$ in a superconductor because the Cooper pairs have zero spin.  Application of the two fluid analysis indicates that the heavy electron component, $K_{HF}(T)$, decreases as expected for a superconducting condensate.  In fact, the data are best fit to a d-wave gap with value $\Delta(0) = 4.5k_BT_c$, in agreement with specific heat measurements. These results strongly suggest that the superconducting condensate arises as an instability of the heavy electron fluid.

% \begin{figure}
%\centering
%\includegraphics[width=\linewidth]{Co115KS.eps}
%\caption{\label{fig:Co115KS} (a) Knight shift of the In(1), In(2) and Co in CeCoIn$_5$ in the paramagnetic state for field in the $ab$ plane. The inset shows the structure of the unit cell, and indicates the position of the different sites and the direction of the applied field.  For In(2) there are two distinct crystallographic positions depending on the direction of the field. (b) The Knight shift in the superconducting state.  The solid points were acquired in a field of 5T and the open points were acquired in a field of 10 T. The solid line in both panels is the susceptibility.  Data are reproduced from \cite{Yang2009}.}
%\end{figure}

 %\begin{figure}
%\centering
%\includegraphics[width=\linewidth]{Co115KSanomaly.eps}
%\caption{\label{fig:Co115KSanomaly} The  heavy electron susceptibility $K_{HF}$ in CeCoIn$_5$ in the normal and superconducting state, measured at 5T, 10T and 12T (??). The solid line in the normal state is the Yang-Pines formula, and the solid line in the superconducting state is computed using a Yosida function (see \cite{Yang2009}).}
%\end{figure}

%\textsc{QUESTION:  Yifeng, these figures are directly copied from our 2009 PRL.  Can we change them somehow?  I think it is important to discuss these results, but only in the context that we already know what happens to the HF fluid in the SC ordered state.}

\section{Hidden order}

Another well known heavy fermion system with long range order at low temperature is URu$_2$Si$_2$, which exhibits a phase transition to a state with an unknown order parameter at $T_{HO} = 17.5$K \cite{MydoshReview}. This unusual phase has been studied for the past two decades and numerous theoretical order parameters have been proposed. Although the nature of the hidden order parameter remains unknown, extensive work has suggested that it has an itinerant nature and involves some type of Fermi surface instability \cite{WiebeURS}. We have measured the Knight shift of the $^{29}$Si in URu$_2$Si$_2$ in order to investigate the effect of hidden order on $K_{HF}$.

Polycrystalline samples of URu$_2$Si$_2$ were synthesized by arc-melting in a gettered argon atmosphere, and an aligned powder was prepared in an epoxy matrix by curing a mixture of powder and epoxy in an external magnetic field of 9 T.  Aligned powder samples are useful to enhance the surface-to-volume ratio and hence the NMR sensitivity. NMR measurements were carried out using a high homogeneity 500MHz (11.7T) Oxford Instruments magnet. The $^{29}$Si($I=1/2$) (natural abundance 4.6\%) spectra were measured by spin echoes, and the signal-to-noise ratio was enhanced by summing several ($\sim 100$) echoes acquired via a Carr-Purcell-Meiboom-Gill pulse sequence \cite{CPSbook}. In this field, the hidden order transition is suppressed to 16 K \cite{JaimeURSPRL2002}. The spin lattice relaxation rate ($T_1^{-1}$) was measured as a function of the angle between the alignment axis and the magnetic field $\mathbf{H}_0$ in order to properly align the sample, since $T_1^{-1}$ is a strong function of orientation with a minimum for $\mathbf{H}_0 \parallel~c$.   Spectra are shown in Fig. \ref{fig:URSmap}, and the Knight shift along the $c$-direction, $K_c$, is shown in Fig. \ref{fig:ShiftURS}. The total magnetic susceptibility, $\chi_c(T)$, is shown in Figs.  \ref{fig:URSmap} and \ref{fig:ShiftURS} as a solid line.  $\chi_c$ exhibited little or no field dependence in the temperature range of interest. The inset of Fig. \ref{fig:ShiftURS} displays  $K_c$ versus $\chi$. There is a clear anomaly that develops around 75 K, in agreement with previous data from Bernal. However, in contrast to previous measurements, we find that below $T^*$ the Knight shift turns upwards rather than downwards \cite{BernalKSAinURS}.

In order to fit the Knight shift data and extract the temperature dependence of $K_{HF}$, we have fit both the $K_c$ versus $\chi$ data and the $K_{HF}$ versus $T$ data simultaneously by minimizing the joint reduced $\chi^2(K_0, B, K_{HF}^0, T^*) = \chi^2_1(K_0, B)+\chi^2_2(K_{HF}^0, T^*)$. Individually, $\chi_1^2$ and $\chi_2^2$ provide the best fits to the two data sets, but since $K_{HF}(T)$ depends on $K_0$ and $B$, $\chi^2_2$ is an implicit function of these fit parameters as well. The best fit is shown in the inset of Fig. \ref{fig:ShiftURS}. The fit parameters, listed in Table \ref{tab:hypIr115}, are close to the values originally reported by Kohori  \cite{kohoriURu2Si2}.   $K_{HF}(T)$ is shown in Fig. \ref{fig:KhfvsT}, and the best fit to Eq. \ref{eqn:YangPines} yields $T^* = 73(5)$ K.  The data clearly indicate that $K_{HF}(T)$ grows monotonically down to the hidden ordering temperature at 16 K and is well described by the Yang-Pines scaling formula.

{Since the $f$-electrons in URu$_2$Si$_2$ are strongly hybridized at low temperatures, the Knight shift and the susceptibility recover a one-component picture below a delocalization temperature $T_L$ \cite{YangPinesPNASdraft}. This allows us to determine the values of $K_0$, $A$ and $B$. The susceptibility of the two components below $T^*$ hence can  be subtracted and the results are plotted in Fig. \ref{fig:URS}. The local component also follows the mean-field prediction for a hybridized spin liquid \cite{YangPinesPNASdraft}, $\chi_l\sim f_l/(T+T_l+f_l\theta)$, with $f_0=1.6$, $\theta=175\,$K given by the RKKY coupling and $T_l=70\,$K by local hybridization.}

Below the hidden order transition temperature $K_{HF}$ decreases dramatically  and saturates at a value $\sim K_{HF}(T_{HO})/3$ at $T\sim 4$ K. This result suggests that, like the superconductivity in CeCoIn$_5$, the hidden order phase in URu$_2$Si$_2$ emerges from the itinerant heavy electron quasiparticles and is not a phenomenon associated with local moment physics \cite{URSmultipole}.  The partial suppression of $K_{HF}$ is also consistent with specific heat measurements which indicate that $\sim40$\% of the Fermi surface remains ungapped below the hidden order transition \cite{MapleURu2Si2}.  Recent theoretical work has suggested the presence of a psuedogap in a range of 5-10 K above the hidden order transition \cite{balatskyURSPG}.  If one neglects the single outlying data point at 16 K, then the data may indicate a subtle change of scaling below $\sim 23$ K.  However we do not have sufficient precision to make any conclusive statements about the pseudogap.

%\textsc{NOTE:  Yifeng, can you do the Yosida function fit to the Knight shift in the HO state and include some discussion here?}

{The Knight shift anomaly $K_{HF}$ below $T_{HO}$ also follows the BCS prediction,
\begin{equation}
K_{anom}(T)-K_{anom}(0)  \sim \int\,dE\left(-\frac{\partial f(E)}{\partial E}\right)N(E),
\end{equation}
where $f(E)$ is the Fermi distribution function and $N(E)=|E|/\sqrt{E^2-\Delta(T)^2}$ is the density of states. The gap function $\Delta(T)$ has a mean-field temperature dependence,
\begin{equation}
\Delta(T)=\Delta(0)\tanh\left[2\sqrt{\left({T_{HO}}/{T}-1\right)}\,\right],
\label{Eq:BCS}
\end{equation}
where $\Delta(0)\sim4.0k_BT_{HO}$ is the gap amplitude determined by STM experiments. We find a good fit to the Kondo liquid susceptibility below $T_{HO}$ in Fig. \ref{fig:URS}.}

\section{Antiferromagnetism and Relocalization}

In order to investigate the response of $K_{HF}$ to long-range order we have measured the Knight shift of the $^{115}$In(1) site in CeRhIn$_5$. CeRhIn$_5$ is an antiferromagnet at ambient pressure with an ordered moment of $0.4\mu_B$ and $T_N = 3.8$K \cite{AnnaCeRhIn5NSpressure}.  Single crystals were grown in In flux using standard flux growth techniques. The $\frac{3}{2}\leftrightarrow\frac{1}{2}$ transition of the $^{115}$In ($I=9/2$) was observed for the In(1) by Hahn echoes,  and the alignment of the crystal was confirmed by the splitting of the quadrupolar satellites \cite{HeggerRh115discovery}.  The Knight shift is shown in Fig. \ref{fig:ShiftRh115}.
There is a clear anomaly that develops at 17 K, in contrast to a previous report of 12 K \cite{CurroKSA}. In this case the previous data consisted of only six data points over a limited range and did not have the precision to discern the anomaly. By taking more data points at closely spaced intervals, we find $K_c$ increases above the extrapolated value below $T^*$, in contrast to that in the CeIrIn$_5$ and CeCoIn$_5$ (see Fig. \ref{fig:115comparison}).  The best fits to the data yield hyperfine couplings that are similar to those in CeIrIn$_5$ (see Table \ref{tab:hypIr115}), and we find that $T^* = 18(1)$ K for CeRhIn$_5$.

The temperature dependence of the emergent heavy fermion component, $K_{HF}(T)$ is shown in Fig. \ref{fig:KhfvsT}(b) for CeRhIn$_5$. In contrast to the scaling behavior observed in CeIrIn$_5$, CeCoIn$_5$ and URu$_2$Si$_2$, $K_{HF}$ \emph{does not} follow the Yang-Pines scaling formula down to the ordering temperature, $T_N$.  Rather, $K_{HF}$ begins to deviate below 9 K, exhibits a maximum at 8K and then decreases down to $T_N$.  This temperature of the breakdown of scaling corresponds well with the onset of antiferromagnetic correlations observed by inelastic neutron scattering and NMR spin lattice relaxation measurements \cite{baoCeRhIn5INS,curroPRL}.  The physical picture that emerges is that the local moments begin to hybridize below $T^*$, but their spectral weight is only gradually transferred to the heavy electron fluid. At 9K, the partially screened local moments begin to interact with one another and spectral weight is transferred back. This \emph{relocalization} behavior was first observed in the antiferromagnet CePt$_2$In$_7$ \cite{CePt2In7KnightShift}.  The fact that $K_{HF}$ remains finite at $T_N$ suggests that the ordered local moments still remain partially screened \cite{YangPinesPNASdraft}.

It is interesting to compare the behavior of the heavy electron Knight shift observed in URu$_2$Si$_2$ and CeRhIn$_5$ with other Kondo lattice materials that undergo long range order. Fig. \ref{fig:scaling} shows the scaled $K_{HF}$ versus $T/T^*$ for URu$_2$Si$_2$, CeRhIn$_5$,  CeCoIn$_5$, and the antiferromagnet CePt$_2$In$_7$ ($T_N = 5.2$ K, $T^* \approx 40$ K) \cite{CurroKSA,CePt2In7KnightShift}. The onset of hybridization at $T^*$ is identical for all the compounds, but the low temperature behavior depends critically on the ground state. In both CeCoIn$_5$ and URu$_2$Si$_2$ the spectral weight appears to be transferred continuously to the heavy electron fluid all the way down to the ordering temperature.  Similar behavior is present in CeIrIn$_5$ ($T_c = 0.4$ K, not shown) down to 2 K. This behavior is surprising, because one might expect that the hidden order and/or superconductivity would emerge from a fully formed heavy electron state.  Such a state would be characterized by a temperature independent $K_{HF}$, but that is clearly not the case for either material.  In fact, the only known cases where $K_{HF}$ appears to plateau at a constant value are CeSn$_3$ and Ce$_3$Bi$_4$Pt$_3$, neither of which exhibit long range order \cite{CurroKSA}.

The relocalization behavior in CeRhIn$_5$ and CePt$_2$In$_7$ contrasts starkly with that of the URu$_2$Si$_2$ and CeCoIn$_5$, where $K_{HF}$ exhibits a peak in the paramagnetic state.  In CePt$_2$In$_7$ $K_{HF}$ was determined via $^{195}$Pt NMR in a powder sample, thus the precision was not as high as for a single crystal in which the full susceptibility tensor can be measured \cite{CePt2In7KnightShift}.  The new results in the CeRhIn$_5$ reported here, however, were indeed acquired in a single crystal and confirm that relocalization behavior appears to be a common feature of antiferromagnetic materials.

\section{Conclusions}

We have measured the emergence of the Kondo liquid and investigated its response to long range order in several heavy fermion compounds.  In the superconductors CeCoIn$_5$ and CeIrIn$_5$ as well as the hidden order compound URu$_2$Si$_2$, $K_{HF}$ rises continuously below $T^*$ down to the ordering temperature without saturation.  In the antiferromagnets CeRhIn$_5$ and CePt$_2$In$_7$ the Kondo liquid relocalizes as a precursor to long range ordere of partially screened local moments.  Much of this behavior can be understood in the the framework of \emph{hybridization effectiveness} introduced in the context of the two-fluid phenomenology \cite{YangPinesPNASdraft}.  In this picture $K_{HF}(T) \sim f_0$, where $f_0$ is a measure of the collective hybridization that reduces the magnitude of the local $f$-moments, and $f_0 = 1$ corresponds to complete hybridization.  CeCoIn$_5$, CeIrIn$_5$ and URu$_2$Si$_2$ correspond to materials with $f_0>1$, whereas CeRhIn$_5$ and CePt$_2$In$_7$ have $f_0<1$. Although the results reported here do not directly address issues of non-Fermi liquid behavior and the presence of quantum critical phenomena in Kondo lattice systems, they do have important implications for theory \cite{SiLocalQCP,ColemanQCreview}.  The dual nature of the coexisting local moments and the heavy electron fluid that exists in these materials below $T^*$ suggests that the physics cannot be described in terms of a single electronic degree of freedom.  Rather, the interactions between the local moments, the heavy electron quasiparticles, and their relative spectral weight should be taken into account.

%% == end of paper:

%% Optional Materials and Methods Section
%% The Materials and Methods section header will be added automatically.

%% Enter any subheads and the Materials and Methods text below.
%\begin{materials}
% Materials text
%\end{materials}

%% Optional Appendix or Appendices
%% \appendix Appendix text...
%% or, for appendix with title, use square brackets:
%% \appendix[Appendix Title]

\begin{acknowledgments}
We thank S. Balatsky, P. Coleman, M. Graf, D. Pines and J. Thompson for stimulating discussions. Work at UC Davis was supported by UCOP-TR01, the National Nuclear Security Administration under the Stewardship Science Academic Alliances program through DOE Research Grant \#DOE DE-FG52-09NA29464, and the National Science Foundation under Grant No. DMR-1005393. Work in China was supported by the Chinese Academy of Sciences and NSF-China (Grant No. 11174339). Los Alamos National Laboratory is operated by Los Alamos National Security, LLC, for the National Nuclear Security Administration of the US Department of Energy under Contract No. DE-AC52-06NA25396.
\end{acknowledgments}

%% PNAS does not support submission of supporting .tex files such as BibTeX.
%% Instead all references must be included in the article .tex document.
%% If you currently use BibTeX, your bibliography is formed because the
%% command \verb+\bibliography{}+ brings the <filename>.bbl file into your
%% .tex document. To conform to PNAS requirements, copy the reference listings
%% from your .bbl file and add them to the article .tex file, using the
%% bibliography environment described above.

%%  Contact pnas@nas.edu if you need assistance with your
%%  bibliography.

% Sample bibliography item in PNAS format:
%% \bibitem{in-text reference} comma-separated author names up to 5,
%% for more than 5 authors use first author last name et al. (year published)
%% article title  {\it Journal Name} volume #: start page-end page.
%% ie,
% \bibitem{Neuhaus} Neuhaus J-M, Sitcher L, Meins F, Jr, Boller T (1991)
% A short C-terminal sequence is necessary and sufficient for the
% targeting of chitinases to the plant vacuole.
% {\it Proc Natl Acad Sci USA} 88:10362-10366.

%% Enter the largest bibliography number in the facing curly brackets
%% following \begin{thebibliography}

%\begin{thebibliography}{}

%\end{thebibliography}
%\bibliographystyle{pnas}
\bibliographystyle{naturemag}

%\bibliography{C:/Users/curro/Documents/Documents/Bibliography/CurroNMR}
\bibliography{C:/Bibliography/CurroNMR}
%\bibliography{C:/Users/Kent Shirer/Documents/UC Davis/Research/My Publications/2012_Knight_Shift}

\end{article}
%%%%%%%%%%%%%%%%%%%%%%%%%%%%%%%%%%%%%%%%%%%%%%%%%%%%%%%%%%%%%%%%

%% Adding Figure and Table References
%% Be sure to add figures and tables after \end{article}
%% and before \end{document}

%% For figures, put the caption below the illustration.
%%
%% \begin{figure}
%% \caption{Almost Sharp Front}\label{afoto}
%% \end{figure}

%% For Tables, put caption above table
%%
%% Table caption should start with a capital letter, continue with lower case
%% and not have a period at the end
%% Using @{\vrule height ?? depth ?? width0pt} in the tabular preamble will
%% keep that much space between every line in the table.

%% \begin{table}
%% \caption{Repeat length of longer allele by age of onset class}
%% \begin{tabular}{@{\vrule height 10.5pt depth4pt  width0pt}lrcccc}
%% table text
%% \end{tabular}
%% \end{table}

%% For two column figures and tables, use the following:

%% \begin{figure*}
%% \caption{Almost Sharp Front}\label{afoto}
%% \end{figure*}

%% \begin{table*}
%% \caption{Repeat length of longer allele by age of onset class}
%% \begin{tabular}{ccc}
%% table text
%% \end{tabular}
%% \end{table*}

%\begin{table*}
%\caption{In(1) Hyperfine coupling constants in CeIrIn$_5$}
%\label{tab:hypIr115}
%\begin{tabular}{@{\vrule height 10.5pt depth4pt  width0pt}lr}
%\begin{tabular}{lr}
%$B_{aa}$  & 11.3(1) kOe/$\mu_B$ \\
%$B_{cc}$  & 13.8(1) kOe/$\mu_B$ \\
%$B_{ac}$  &  0.4(3) kOe/$\mu_B$ \\
%$K^0_{aa}$  & 1.1(2) \% \\
%$K^0_{cc}$  &  0.4(2) \% \\
%$K^0_{ac}$  & -0.5(2) \%
%\end{tabular}
%\end{table*}

\begin{table*}
\caption{Crystal Field Levels and Coherence Temperatures in the CeMIn$_5$ Materials}
\label{tab:CEF115}
\begin{tabular}{@{\vrule height 10.5pt depth4pt  width0pt}lrr}
%\begin{tabular}{lr}
Material & $T^*$ (K) & $\Delta$ (K)(from \cite{CEF115study})\\
CeCoIn$_5$ & 42 & 100\\
CeRhIn$_5$ & 18(1) &80 \\
CeIrIn$_5$ & 31(5) &78 \\
\end{tabular}
\end{table*}

\begin{table*}
\caption{Measured hyperfine coupling constants in CeIrIn$_5$, CeRhIn$_5$ and URu$_2$Si$_2$}
\label{tab:hypIr115}
\begin{tabular}{@{\vrule height 10.5pt depth4pt  width0pt}llrrrrrr}
%\begin{tabular}{lr}
Material & site & $B_{aa}$ (kOe/$\mu_B$)  & $B_{cc}$ (kOe/$\mu_B$)  & $B_{ac}$ (kOe/$\mu_B$)  & $K^0_{aa}$ (\%) & $K^0_{cc}$ (\%) & $K^0_{ac}$ (\%) \\
CeIrIn$_5$ &In(1) & 11.3(1)   & 13.8(1)  & 0.4(3)  &  1.1(2)  &  0.4(2)  &  -0.5(2)\\
CeRhIn$_5$ &In(1) & --   & 21.4(5)  & --  &  --  &  0.64(5)  &  --\\
URu$_2$Si$_2$ &Si & --   & 4.3(1)  & --  &  --  &  -0.07(1)  &  --
\end{tabular}
\end{table*}

\end{document}